# Industrially Applicable System Regression Test Prioritization in Production Automation

Sebastian Ulewicz and Birgit Vogel-Heuser, *Senior Member, IEEE*

*Abstract*—When changes are performed on an automated production system (aPS), new faults can be accidentally introduced into the system, which are called regressions. A common method for finding these faults is regression testing. In most cases, this regression testing process is performed under high time pressure and on-site in a very uncomfortable environment. Until now, there has been no automated support for finding and prioritizing system test cases regarding the fully integrated aPS that are suitable for finding regressions. Thus, the testing technician has to rely on personal intuition and experience, possibly choosing an inappropriate order of test cases, finding regressions at a very late stage of the test run. Using a suitable prioritization, this iterative process of finding and fixing regressions can be streamlined and a lot of time can be saved by executing test cases likely to identify new regressions earlier. Thus, an approach is presented in this paper that uses previously acquired runtime data from past test executions and performs a change identification and impact analysis to prioritize test cases that have a high probability to unveil regressions caused by side effects of a system change. The approach was developed in cooperation with reputable industrial partners active in the field of aPS engineering, ensuring a development in line with industrial requirements. An industrial case study and an expert evaluation were performed, showing promising results.

*Note to Practitioners*—Currently, prioritizing relevant system tests to be executed in case of changes to an automated production system (aPS) is very challenging and depends largely on the intuition and experience of the testing technician. As system tests involve manual operations, testing requires substantial effort. In addition, testing is mostly performed under severe time pressure and in an uncomfortable environment such as on-site at the customer's premises. In this work, an approach is presented that supports the testing technician in finding and prioritizing available test cases based on previous test executions and change analysis. For practitioners, this approach could streamline regression testing and increase the code quality of aPSs significantly, in particular, for fully integrated aPSs, which have not been the focus of other research so far. As the approach was developed in line with industrial requirements, relevance and adaptability for industrial use are ensured.

This paragraph of the first footnote will contain the date on which you submitted your paper for review.

Substantial parts of the approach presented in this paper were supported by the Bavarian Ministry of Economic Affairs and Media, Energy and Technology within the research program *Information technology and communication technology in Bavaria* (Germany) under Grant IUK413.

The authors are with the Technical University of Munich, Chair of Automation and Information Systems, Boltzmannstr. 15, 85748 Garching, Germany (e-mail: {sebastian.ulewicz; vogel-heuser}@tum.de).

## I. INTRODUCTION

AUTOMATED production systems (aPSs) have high demands regarding reliability and availability [1], as interruptions in production or products of insufficient quality result in great financial losses. As the complexity of these systems is constantly rising, particularly their control software [2], quality assurance is of high importance. Even though model-driven engineering methods [3], component architectures [4] and approaches for distributed systems [5] have been proposed in research to manage the program complexity, most industrial aPSs are still directly programmed in the standard IEC 61131-3 [6]. To ensure a high level of system quality, typically system tests are performed to find faults, e.g. unwanted behavior. Commissioning or maintenance personnel, which requires substantial amounts of human resources and time, typically perform these tests manually until no new faults are identified during the testing process.

APSs are often subject to changes after the start of production, e.g. due to changed requirements by the customer, newly found bugs in the control software or wear on hardware components. During these modifications, new faults can unintentionally be implemented. To find these so-called regressions of the system, regression testing is performed. During this process, previously successfully performed test cases are performed again to identify possible unwanted side effects of changes. During this process, the involved personnel are under high time pressure, as regular operation and production are halted, causing substantial financial losses. Thus, the testing process is to be kept as short as possible, while assuring sufficient system quality. The difficulty and problem for the involved personnel is to identify and perform only relevant test cases for the implemented change under the difficult situation of high time pressure without any support by automated systems. This problem is exacerbated by the necessity to restart the regression testing process after fixing a regression.

In this work, an approach will be presented that provides the personnel involved in the system regression testing process with automated support in prioritizing test cases based on the implemented change of the system to allow for quick identification of newly introduced unwanted behavior. The approach is based on a guided semi-automatic system testing approach [7], which allows for structured system tests on fully integrated aPSs, integrating a human operator. This approach was extended by a method to build a relation between each system test and the related parts of the control software of the aPS by using so-called execution tracing. After changes to the system, this information is combined with a change identification and change impact analysis to allow for a prioritization of test cases with a high probability to identify newly introduced faults. For this, test cases related to changed and possibly influenced parts of the code are prioritized before the rest of the test cases. The approach was developed in close cooperation with industry, taking important requirements in this domain into account, particularly the absence of detailed simulations or formal specifications and the need for real-time capability. To allow for an assessment of the industrial applicability of the approach, a case study using a real aPS was conducted. The results of this case study were discussed with industrial experts, allowing for a qualitative evaluation of the approach. The approach is an extension and improvement of the work presented in [8], where a basic tracing and prioritization method using a laboratory plant was shown. This approach was extended by a change impact analysis, significantly refined regarding the prioritization and optimized for industrial application by performing an industrial case study and expert evaluation.

The remainder of this paper is structured as follows: Industrial requirements for the approach are presented in section II, after which related work and the research gap are discussed in section III. In section IV, the concept is described in detail. Section V provides a brief explanation of the implementation used in the evaluation presented in section VI. Following the evaluation, several performance improvements of the approach are presented in section VII. The last section gives a conclusion of the approach and an outlook on future research.

## II. Industrial Requirements Regarding System and Regression Testing in Production Automation

The presented approach was developed in close cooperation with industrial experts. In several workshops with up to seven experienced experts from three different internationally renowned companies related to or active in the field of factory automation, requirements were specified, which are presented in the following. The scenario for which these requirements were developed is the testing process of fully integrated industrial aPSs, in particular, the regression testing process performed after a tested system is modified.

*R1 Support of industrial aPS software properties*: The approach is to be applicable for industrial standards of control software programming, in particular for the IEC 61131-3 programming standard.

*R2 Real time capability and memory size*: The approach is not to influence the control software in a way that would not allow the real time requirements of the system to hold with the currently set maximum scan cycle time.

*R3 Inclusion of valid hardware and process behavior in the testing process*: The approach is to test the system including valid (real) behavior of the hardware and the technical process.

*R4 Manipulation of hardware and process behavior during testing*: The approach is to allow for the manipulation of the hardware and process behavior during the testing process.

*R5 Increase in efficiency during the testing process of changes to a previously tested control software*: The approach is to improve the efficiency during the testing process required after the implementation of software changes to a previously tested system that do not relate to changed requirements, e.g. bug fixes or optimizations.

Based on these requirements, related work in the field of automated production systems and adjacent domains was analyzed for applicability and a research gap was identified. Subsequently, the requirements were the foundation for developing and evaluating the presented approach.

## III. Related Work

To identify the research gap the approach is to fill, related work in the field of quality assurance and test prioritization methods in the aPS domain were analyzed and rated for their applicability.

A multitude of approaches was developed for improving the quality assurance of software in general and control software of aPSs in particular. Besides static approaches regarding the structure of control software [9] or integrated system models [10], several dynamic testing or verification approaches can be identified. These approaches can be clustered into the fields of formal verification, model-based test generation and virtual commissioning.

Formal verification is used to mathematically prove the compliance of a model to a specification. Different approaches and tools were developed for aPS, such as the formal verification tool Arcade.PLC [11] or a specification language and verification tool for individual software components [12]. While a system's compliance with the specification can be exhaustively proven, the approach requires extensive resources for the specification of formal requirements and system models in addition to an extensive knowledge of formal methods and often fails to deal with the complexity of fully integrated systems ("State Space Explosion"). Other approaches try to mitigate the problem of resource intensive specification efforts and the complexity problem by focusing on performed changes [13]. Yet, disregard for valid hardware behavior remains.

Using model-based test generation and test automation techniques [14], the effort for specifying and performing test cases can be reduced. Formalized functional specifications can be used to describe the intended system behavior [15] or fault handling functionality of the system [16]. While the test specification effort can be reduced, the approaches require simulations of hardware and technical process behavior or do not take testing after changes into account.

Virtual commissioning techniques have proven to be valuable for aPSs produced in greater lot sizes. For this, detailed simulations are specified, allowing for an automatic execution of a multitude of test scenarios [17]. Unfortunately, the creation of simulations requires extensive effort to allow the representation of valid system behavior regarding the hardware and technical process. This problem can be mitigated by designing simulations of different abstraction levels related to the tested problem [18] or by using existing engineering artifacts for an automatic generation of simulation models [19]. In many cases, the required documents and resources for the creation of the simulations are not available in industry, especially for individually engineered machines.

Besides the general approach to make the testing of aPSs more efficient and increase testing quality as presented in the previous paragraphs, several approaches focus on the selection and prioritization of existing test cases. In the domain of computer science in particular, this has been an active field of research for many years [20]. Two main classes of approaches can be identified in this field: static and dynamic techniques. Static techniques focus on the connection of engineering artifacts to the test cases (traceability) to gain information about what test case should be selected and prioritized due to a change. Dynamic techniques leverage data acquired during the execution of test cases to allow for a relation of the test cases to the tested code. After changes to the system, this allows for an assessment of test cases with a higher probability of yielding different results, e.g. by failing, exposing newly introduced unwanted behavior. While static analysis generally is better at finding all relevant test cases (soundness), dynamic approaches in computer science often choose less unnecessary test cases [21] and are thus more precise.

Regarding static traceability methods, established tools for requirements engineering can be used to infer connections between requirements and test cases [22]. In case of changed requirements, related test cases can quickly be identified. Yet

TABLE I
OVERVIEW OF RELATED APPROACHES OF SYSTEM REGRESSION TESTING AND RATING USING INDUSTRIAL REQUIREMENTS FOR APS

| Approaches | R1 | R2 | R3 | R4 | R5 |
|---|---|---|---|---|---|
| Biallas et al. [11] | + | + | - | - | ○ |
| Ljungkrantz et al. [12] | + | + | - | - | ○ |
| Ulewicz et al. [13] | + | + | - | - | + |
| Hametner et al. [15] | + | + | ○ | + | - |
| Rösch & Vogel-Heuser [16] | + | + | + | + | - |
| Süß et al. [17] | + | + | ○ | + | ○ |
| Puntel-Schmidt & Fay [18] | + | + | ○ | + | ○ |
| Barth & Fay [19] | + | + | ○ | + | ○ |
| Req. management tools [22] | + | + | + | + | - |
| Zeller & Weyrich [23] | + | + | ○ | + | + |
| Caliebe et al. [24] | ○ | + | - | - | + |
| Baller et al. [25] | ○ | + | - | - | + |
| Ulewicz et al. [26] | + | + | - | - | ○ |
| Orso et al. [27] | - | - | - | - | + |
| Rothermel et al. [28] | - | - | - | - | + |
| Prähofer et al. [29] | + | + | + | + | - |

+: fulfilled, ○: partially fulfilled, -: not fulfilled

R1: Support of industrial aPS software properties, R2: Real-time capability and memory size, R3: Inclusion of valid hardware and process behavior in the testing process, R4: Manipulation of hardware and process behavior, R5: Increase in efficiency during the testing process of changes to a previously tested control software in practice, changes are often directly performed on the system, often without a change of the requirements' documents. In these cases, a detailed relation of test cases to the performed change is impossible to achieve, hindering a prioritization of available test cases. Several approaches try to improve this by including additional information, such as historical data for decentralized production systems [23] or system models [24]. For product lines, test-case selection can also be based on a product line model [25]. Still, all of these approaches require additional artifacts and their connection to the system, which are often not available in industry. One approach for test selection is independent from additional models, solely basing the choice of test cases on changes to the software [26], yet the approach is limited to unit and integration testing, and is not applicable to system testing.

When using dynamic methods, traces are recorded during the execution of test cases to allow for a relation between executed code and its related test case. By identifying changes to a system, this information can be used to prioritize available test cases [27], [28]. As these works stem from computer science, important requirements for aPSs are not regarded. In particular, the test cases used in the approaches are performed completely automatically, with no possibility for manual interaction with the system, and their runtime overhead is not investigated. Only one approach for tracing seems applicable for aPSs, yet the approach focuses on reproducing variable values for manual debugging and does not consider test automation or regression testing [29].

As the discussion of the related work in the previous sections shows, none of the analyzed approaches succeed in fulfilling the imposed requirements identified in the workshops with the industry partners (section II). As compiled in Table I, this is mostly due to not taking valid hardware behavior into account sufficiently. In addition, the disregard of important real-time requirements and industrially relevant test scenarios, including manual interaction with the system, prevents all approaches from computer science from being directly applicable. Thus, the identified research gap and contribution of this work is a test prioritization method aimed at system tests for aPSs, which is in line with the prevalent requirements. In particular, the approach takes hardware and process behavior into account, while remaining independent from simulations and taking real-time restrictions into account. The aim of this prioritization is to identify regressions in the system earlier, optimizing the debugging process after implementing changes to a previously tested system.

## IV. CONCEPT

The presented approach focuses on a prioritization of functional black box tests [30] on the system level for a previously tested system that has undergone changes. Using a previously presented concept [7], these tests are semi-automatically executed using a fully commissioned aPS and an operator (see Fig. 1, top right). During test execution, instructions are given to and acknowledged by the operator using a human machine interface (HMI). This is achieved by including test cases and an additional tracing functionality in the original control program of the aPS (see Fig. 1, top left).

While this previous work only focused on the test execution itself, the prioritization of such test cases is the focus of this work.

The goal of the prioritization is to efficiently find newly introduced faults in a previously tested aPS that has now been modified, using existing test cases. To enable this efficient regression testing process, the system tests to be executed are arranged in an optimized order: test cases with a higher probability to find newly introduced faults are moved to the front of the queue. Through this, possible regressions of the system can be found earlier, allowing for an optimized iterative debugging process (fixing the regression and repeating the regression testing process).

As schematically shown in Fig. 1 (bottom), the approach comprises three steps, which succeed the preparation (top left) and testing of the previous version of the aPS (top right):

Step 1: A relation between each test case and the executed control program parts is built for the previously tested version of the aPS. This is based on test results and execution traces acquired during test execution on the previous aPS version.

Step 2: Changes between the previous and current version of the control program and possible impacts of these changes are identified.

Step 3: Previously existing system tests for the current version are prioritized. This is based on possible impacts of the changes (step 2) and runtime and timing information acquired during previous test executions (step 1).

Building a relation between code and system test cases is not trivial compared to unit tests, where test cases are directly related to specific components of a system. While unit tests usually set and check most of the units' variable values very specifically, system tests use fewer, high-level stimulations and checks to initiate and perform tests. For example, rather than exactly specifying all sensor and expected actuator values

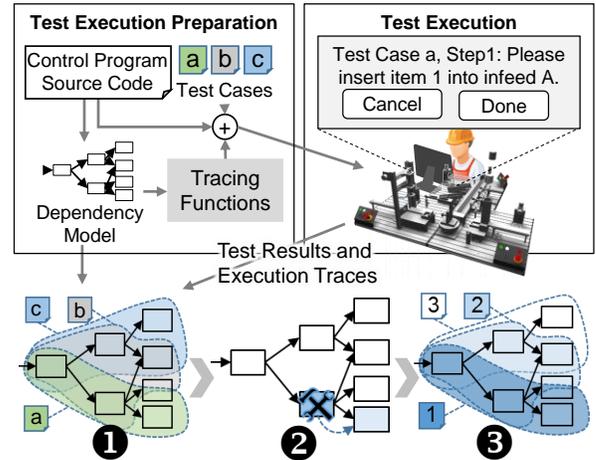

Fig. 1. Overview of the approach: After an original control program is extended by test cases and tracing functions (top left) it can be used for semi-automatic system testing (top right). Based on the results and execution traces recorded during the testing process, a relation between system tests and code is built (1), changes and possibly influenced parts of the original code are identified (2) and system tests are prioritized for a changed aPS regarding their likelihood to unveil new faults (3)

for a certain function, the system is stimulated by starting an automated process via the HMI and inserting an intermediate product into the machine. Which exact parts of the code relate to the tested function is not defined and remains unclear unless a relation is being built, as performed in step 1. This information can then be used to focus the change impact analysis and prioritization in steps 2 and 3, which would struggle finding related system test cases if only static methods were used.

The foundation of the approach is a dependency model that is used to identify suitable points in the control code of programmable logic controllers (PLCs) to acquire runtime information during test execution and to identify changes in the PLC control program. In the following sections, this dependency model will be presented, followed by each of the steps of the prioritization approach.

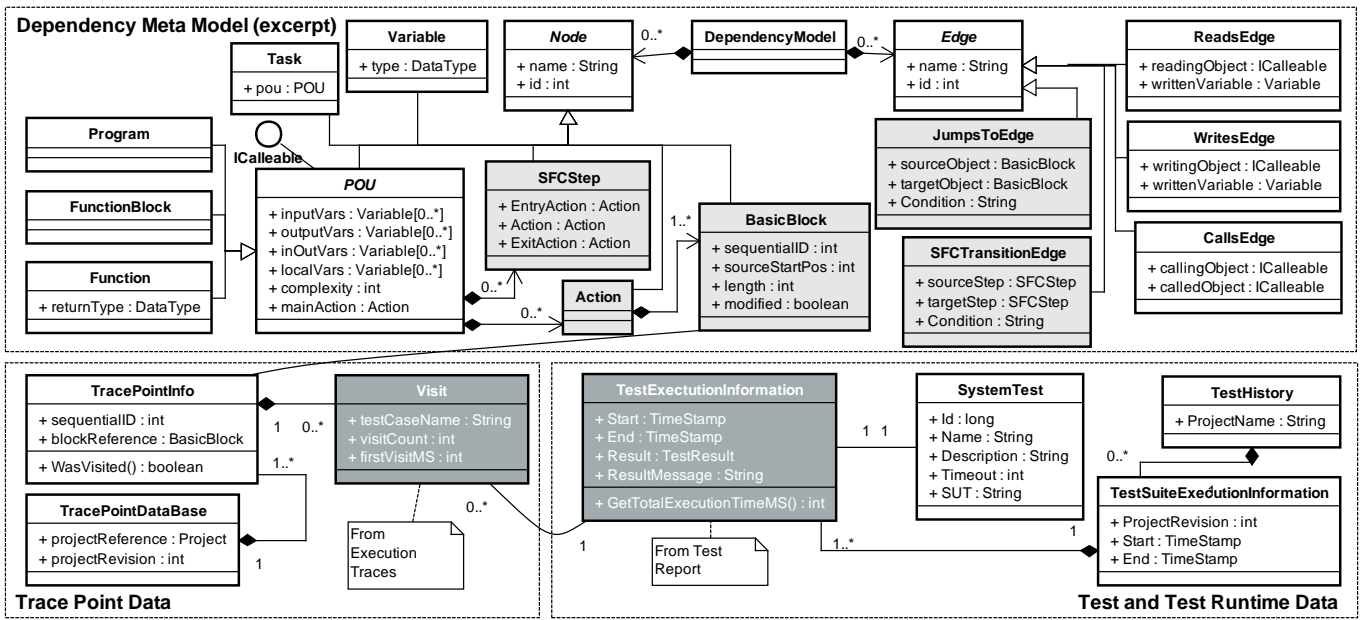

Fig. 2. Overview of the used meta models and their connections: the dependency model (top) extended from [31] (extension in light grey), the trace-point data (bottom left) and the test and test runtime data (bottom right). Connection between code (basic blocks) and system tests is achieved through trace-point visits (recorded during test execution) directly related to basic blocks.

*A. PLC Control Programs and the Dependency Model*

PLC control programs programmed in IEC 61131-3 [6] consist of program organization units (POUs) that contain executable code implemented in one of five different programming languages defined in the IEC 61131-3, and globally defined variables and data types, among other entities. Depending on the task configuration, PLC programs are almost exclusively executed cyclically, meaning that a standard PLC scan cycle is performed: reading sensor inputs, executing the program starting from the POU(s) specified in the task configuration and writing actuator outputs. This scan cycle is repeatedly executed, with each repetition lasting no longer than a specified maximum PLC scan cycle time for each task to satisfy real-time requirements (typically in the range of 1-100 milliseconds for logic tasks). During the execution of the POUs, the program execution can progress through different paths in the code, called the control flow, based on the given inputs and current values of internal variables.

For modularity and maintainability analysis based on the control and data flow within control programs, a dependency model definition was presented in previous work [31] that is used to generate a dependency model from a PLC control software project, in contrast to other approaches using model-driven engineering to generate PLC control software [3]. The related meta model, which is partially presented in the upper part of Fig. 2, was extended by stereotypes to include control flows within POUs (extensions in light grey). The extended dependency model is a directed graph comprising nodes and edges. Nodes represent structural entities of an IEC 61131-3 project, whereas edges represent the dependencies between these entities. An edge is a unidirectional connection between two nodes: a source and a target. The extended meta model defines nodes from a control software project such as tasks, POUs (functions, function blocks and functions), sequential function chart (SFC) steps and transition, POU actions and basic blocks (code segments that contain no decisions). Dependencies are represented by edges between nodes, such as calls between POUs (functions and programs) or POU instances (function blocks), read and write operations and progressions between basic blocks (JumpsToEdge) and SFC steps (SFCTransitionEdge). So far, the dependency model supports the detailed description of IEC 61131-3 structured text (ST) and SFC implementations.

*B. Acquiring System Test Runtime Information*

A substantial input for the prioritization algorithm is test results and data acquired during runtime for each test case. The test results include information such as the test verdict (passed/failed) and total execution time. The additionally acquired runtime information for each test case is comprised of (i) what parts of the code were active during its execution, (ii) the number of times the test case passed through each part of the code and (iii) when it passed through each part for the first time. As a foundation for the acquisition of this runtime data, the code is instrumented: the original project is automatically extended by functions and function calls. This instrumentation is used during the execution of each system test to record runtime data using a guided semi-automatic system testing approach [7].

*1) Code Instrumentation for the Acquisition of Relevant Runtime Information*

The instrumentation of the code is the foundation of the acquisition of runtime information during test execution. The instrumentation consists of creating structures designated for the recording of trace information and the inclusion of tracing statements into the control program at relevant points in the source code. The insertion of the trace statements (e.g. "tp.x*25* := TRUE;" for tracing code traversals, where 25 is the trace point ID) is described in Fig. 3. The previous version of the tracing approach is explained in more context and detail in [8]. This previous approach was extended by the possibility to record how often each relevant part of the code was traversed by each test case and when it was traversed for the first time. In addition, for performance reasons, previously used call-by-value function calls for tracing ("tpr(i:=*25*);", 25 being the trace point ID) were exchanged for inline tracing statements, and trace arrays for storing tracing information were replaced by structures. These improvements were performed after the initial expert evaluation (see section VI), thus the runtime properties with this current state of the approach yield better results (see section VII). To this point, this tracing approach allows for the recording of whether, how often and when a part of the code was executed for the first time, yet does not distinguish by whom the traversal of this part was initiated. Thus, for a multi-task PLC program, no distinction of calling tasks is made, as this is not required for the current state of the approach.

*2) Recording and Saving Runtime Information Using a Guided Semi-Automatic System Testing Approach*

Currently in industry, system testing is almost exclusively performed manually and does not allow any recording of runtime information regarding test coverage. For this reason, the presented approach is based on a semi-automatic guided

```
1.  def nodes = set of all nodes in dep. model
2.  def instrumentables = nodes.FindAll(where type
      is Function, Program, FunctionBlock or Action)
3.  def tpCounter = 0  //counter trace point index
4.
5.  for each instrumentable in instrumentables
6.    def basicBlocks = nodes.FindAll(where type
        is BasicBlock and parent is instrumentable)
7.    sort basicBlocks by sourceStartPos
8.    offset = 0  //source pos. for instrum.
9.    for each basicBlock in basicBlocks
10.     text = "tp.x" + (tpCounter) + " := TRUE;"
11.     position = basicBlock.sourceStartPos +
          offset
12.     instrumentable.insertText(text, position)
13.     offset += text.length
14.     tracePointDataBase.AddTracePoint(
          tpCounter++, basicBlock.ID)
```

Fig. 3. Pseudo code describing the instrumentation process: A trace function (line 10) is added in front of every basic block; information about this inserted code is stored in the trace-point database for later utilization

system testing approach previously published in [7]. Here, each system test can comprise automatic stimulations and checks of the control program and manual tasks given to a testing technician (ranging from inexperienced workers to experienced engineers). These tasks are given using an HMI, such as a display and touch interface as currently available in most aPSs. Using this method, system tests become more structured and allow for the automatic recording of runtime information, among other positive aspects, e.g. a more structured test execution and automatic detailed documentation of the testing process.

Each test case is extended by calls of the tracing functions "reset" at the beginning of each test case, which resets trace structures that store execution trace information, and "save" at the end of each test case, which saves the information in the trace structure to the PLC hard drive. During test execution, the trace structure is filled with information about traversals and timing information upon executing the instrumented control code. After the execution of all test cases, the saved test traces are transferred from the execution hardware (PLC or Embedded PC) to the engineering platform (PC with IDE) for analysis and storage.

*3) Relating Code to System Test Cases*

Using the information acquired during test execution and instrumentation, a relation between the tested code (basic blocks) and the individual system test cases can be inferred. The meta model for storing this information is shown in Fig. 2, bottom left. During instrumentation, the trace-point database is generated, relating basic blocks to the individual trace points. After test execution, the test report, which stores information about which test cases were executed and information about their success and required total execution time, is passed and stored in the respective test execution information object. By parsing the individual execution traces generated during test execution, information about individual test cases "visiting" (traversing) each trace point, including information about the number of visits and their first visit, can be extracted. Using this information, a direct relation between each test case and the basic blocks can be generated. Upon change of the system, this information is used to extract the required information for prioritization.

*C. Acquiring Change Impact Information*

Regression testing uses previously successfully executed test cases to possibly identify newly introduced faults in the system caused by a change. For this approach, it is assumed that all test cases for which the prioritization is to be performed are still technically executable (e.g. no changes of variable names checked by test cases) and valid (e.g. no change of a requirement making a test case inconsistent with the requirement). Changes that would require an adaption of existing test cases are not regarded in the current state of the approach. In addition, newly specified test cases are also not part of the prioritization for regression testing but should be executed first to test the change itself.

*1) Software Change Identification*

Changes in a software can be identified by comparing a previous (unchanged) and a current (changed) software revision. Current Integrated Development Environments (IDEs) often already offer syntactic change analysis of control software, but lack the identification of changes of the control and data flow. As the presented regression test prioritization is based on the relationship between test cases and the program control flow, the comparison is directly performed with the revisions of the system's dependency model. Changes are identified in a top-down manner: coarse-grained changes are identified by comparing the items of the dependency model and, subsequently, fine-grained changes are identified by comparing modified items in more detail.

*Coarse-grained changes (software project level)*: Through comparison of the set of nodes and edges of the project level (POUs and calls), added, removed and modified nodes and edges can be identified. The relation between the items in the sets of old and new are generated by the items' qualified name (unique name based on their parent objects' names and own name). Items of the same name that have a changed checksum of their content are marked as "modified" (fine-grained changes) and are subsequently analyzed in more detail by regarding their sub-items. In addition, modifications on globally specified variables are analyzed in a similar way.

*Fine-grained changes (source code level)*: Modified POUs are analyzed for their internal changes by comparing their control flow with the unchanged revision. Depending on the implementation of the POU, this is directly performed on the basic blocks and decisions stemming from an ST implementation or the SFC steps and transitions of an SFC implementation. Changes to SFC elements are identified similarly to coarse-grained changes: the set of SFC steps is compared using the steps' names, followed by a comparison of the connecting transitions. Actions related to the SFC steps are then compared, similar to ST implementations. Small changes in the ST control flow, i.e. changes that only modify existing basic blocks or transition conditions, are identified by comparing the control flow graphs. In case transitions were modified, the previous and following basic blocks are marked as modified, as the tracing algorithm focuses on the traversal of basic blocks. Larger changes to the control flow, i.e. changes that add and remove basic blocks or transitions, cannot be safely identified by the comparison algorithm. This is a focus of future work. In case the comparison algorithm fails to identify small changes, all basic blocks are marked as changed to be sure all test cases relating to this POU or SFC step will be prioritized.

*2) Software Change Impact Analysis*

A change to the control software cannot only have direct influence on the software's output signals and their timing, but also on other elements in the code. If, for example, a local variable is assigned a different value in a modified basic block, this can have an influence on the program's progression through the control flow. Regarding this indirect influence on the control software's behavior, a change impact analysis algorithm was developed to analyze the cross connections within the program based on the previously identified changes. For this, modified basic blocks, transitions between basic

```
1.  def modBBs = nodes.FindAll(where type is
    BasicBlock and modified is true)
2.  for each modBB in modBBs
3.      //Get added, changed or removed statements
4.      m_sts = modBB.GetAllModifiedStatements()
5.      for each m_st in m_sts
6.          analyzeStatement(m_st)
7.
8.  function analyzeStatement(statement st)
9.      if st is decision
10.         st.SetSourceBasicBlockModified()
11.         st.SetTargetBasicBlocksModified()
12.     else
13.         st.SetParentBasicBlockModified()
14.     if st is assignment //variable := value
15.         analyzeStatement(st.variable)
16.     if st is call //callee(param. assign.)
17.         for each pa in st.parameterAssignments
18.             analyzeStatement(pa)
```

Fig. 4. Pseudo code describing the change impact analysis: each modified basic block is analyzed for modified statements, which are subsequently analyzed for their impact on other basic blocks

blocks and SFC steps and global variable assignment are analyzed in detail for three different types of possible influences: influence thorough changed assignments, calls or decisions (see Fig. 4). Directly changed or indirectly changed items both will be called "modified" from this point on for better readability.

*Influence by modified assignments*: A change of an assignment of a variable (write access) that is used in a different part of the code (read access) can have an influence on the progression through the code (control flow) or the output behavior of the control program. Thus, the newly assigned variable is marked as modified.

*Influence by modified calls*: A change in passed values in a POU (instance) call will likely have an influence on its behavior. Therefore, the called POU is marked as modified in case of modified passed values in one or more arguments.

*Influence on the control flow by modified decisions (edges between basic blocks)*: If a condition of a decision was modified by another change (see previous change types), the progression of the program through the code (control flow) is likely to exhibit differing behavior. As decisions are not instrumented directly, the previous and subsequent basic blocks of the decision are marked as modified.

Using this change impact algorithm, the influence of a change can be analyzed and is stored in the dependency model: basic blocks are marked as modified, which is treated equivalently to a direct change of a basic block. The change can affect the control flow within a POU, but can also reach code outside of a POU. While the influence could technically span throughout the whole program quickly, this problem was not encountered in any of the conducted preliminary and evaluation experiments. Still, optimizations of this algorithm can be the focus of future research.

*D. Test Prioritization*

The system test-case prioritization is based on the relation of each test case to the control program, test-case timing information and modifications (changes and possible change influences) to a system's control software.

The test cases are prioritized in two steps. In the first step, each test case traversing modified parts of the code is prioritized. In the second step, the prioritization is refined using one of two strategies depending on the user's choice: prioritizing test cases that intensely or quickly traversing all modifications. In the following sections, these prioritization methods will be presented in detail.

*1) Simple Prioritization: Prioritizing Test Cases That Traverse Modifications*

The goal of a prioritization is to quickly unveil regressions in the system. Test cases that directly traverse modifications are more likely to unveil regressions by failing than others that are executing completely unchanged code. For this reason, the set of system tests is grouped into test cases that are modification-traversing (high-priority test cases) and those that are not (low-priority test cases). The grouping is performed in four steps: I) all possibly influenced basic blocks are identified using the dependency model, which stores information about changes and possible influence (modifications) after the change identification and change impact analysis. II) An iteration through each execution trace is performed and "visited" (traversed) basic blocks that were possibly influenced are identified (compare Fig. 2, "visits" on bottom left). III) If an execution trace shows that a test case has visited a modified basic block, it is added to the group of high-priority test cases. IV) After iteration through all execution traces, all remaining test cases are added to the group of low-priority test cases.

In practice, systems are tested with a set of many test cases. Thus, many test cases might be identified as high-priority test cases. For a more efficient prioritization, a refined prioritization was developed.

*2) Refined Prioritization I: Prioritizing Test Cases That Intensely Traverse Modifications*

Some changes of the system cause regressions that might not become apparent in the first traversal of modified code. This more detailed prioritization, therefore, gives test cases a higher priority that traverse modifications as much as possible in the least amount of time. It is aimed at finding sporadic faults caused by regressions of the system.

The information that is gathered for this prioritization is the amount of traversals of possibly influenced parts of the code and the total execution time previously required by the test case. For each test case, a prioritization number $p_{it}$ is calculated. This number represents the times per second the test case previously traversed now-modified basic blocks. The number is calculated as follows: $p_{it}$ is the sum of all traversals of all modified basic blocks divided by the seconds previously needed to execute the test case.

After calculation of the prioritization number $p_{it}$, a refined prioritization of all test cases can be performed. So far, the algorithm does not differentiate between different modifications. Thus, if a change has an impact on many parts of the code, this prioritization algorithm might not check the desired part of the functionality. This prioritization method is therefore aimed at changes that have little influence on the control code, but might fail due to sporadic faults, in particular, in connection with the controlled hardware.

*3) Refined Prioritization II: Prioritizing Test Cases That Traverse as Many Modifications as Fast as Possible*

Some changes influence many different parts of the code that might all be related to regressions of the system. For this reason, this refined prioritization method tries to prioritize a set of test cases that executes as many modifications of the code as fast as possible. It is aimed at quickly finding faults that become apparent in the first traversal of the modified part of the system.

If a revision of the control software includes many modifications (changes to the control software and resulting possibly influenced parts of the code), there might not be a single test case that traverses all of these modifications. Thus, instead of prioritizing single test cases, modification-traversing test combinations (MTTCs) are arranged. These combinations are sequences of test cases to be executed to cover all or as many modifications as possible. The MTTCs are inferred using the following process:

1. For each modification-traversing test case, a new MTTC is instantiated.

2. If the test case does not cover all modifications yet, all test cases related to the remaining untraversed modifications are combined with it. Each combination results in a new MTTC.

3. For each MTTC, the total time needed to traverse all modifications is calculated.

4. Each MTTC instance is then rated by the needed total time to traverse all modifications. The fastest MTTCs are prioritized highest.

5. After finding the fastest MTTC, this process is iteratively repeated on the remaining test cases.

Through this prioritization, a combination of test cases that previously traversed many or all possibly influenced parts of the control software are executed first, followed by further combinations that try covering as many modifications as possible. Non-modification-traversing test cases follow as low-priority test cases with no particular order. Using this method, a quick test of all possibly influenced parts of the code is achieved, allowing one to unveil potentially introduced unwanted behavior in different functions of the code.

## V. IMPLEMENTATION IN AN INDUSTRIAL PROGRAMMING ENVIRONMENT

To allow for an evaluation of the applicability of the presented concept, a prototypical tool was implemented. This tool enables the definition and execution of system tests, their subsequent prioritization and the measurement of their execution time properties. For this, a plug-in for the widely used CODESYS V3.5 Integrated Development Environment (IDE) for Automated Production Systems programmed in the IEC 61131-3 standard [32] was implemented. Building on the close integration with the IDE, an almost fully automated workflow and algorithms using reliable information provided by the IDE were implemented into the plug-in. Test definition and test project generation were adapted from the previously developed semi-automatic system testing approach [7]. It is built upon the CODESYS Test Manager [33] extended by invoking tracing functions (see section IV.B) to record runtime information for each test case. The resulting execution traces were stored directly on the embedded PC during test execution. The traces are automatically transferred to the development system for subsequent analysis after all tests are executed. For this, the developed plug-in automatically loads and analyzes the execution trace files and calculates the prioritization of the test cases according to the chosen prioritization method. This implementation was based on a previous implementation [8] that was extended by the different prioritization methods and refined regarding performance and applicability for industrial use. For experimentation purposes, the instrumentation algorithm was implemented in such a way that either no or a specific instrumentation was possible (depending on the used prioritization algorithm) to acquire more information about the differences in the needed execution time and the memory overhead.

## VI. EVALUATION

Several experiments were designed in cooperation with industrial experts and performed by the authors for the evaluation of the approach. As proposed by [34], a representative group of participants and data measurements were intentionally chosen to allow for an evaluation of the initially defined requirements. The measurements include execution timing to assess the approach's influence on runtime properties and thus its applicability for the production automation domain. Additionally, the prioritization results were recorded. The experiments' results were discussed in a workshop with six experienced experts from the field of aPS engineering from one of the participating companies.

The following sections present the case study, a decommissioned machine formerly used in industry, the experiments and performed measurements. Subsequently, the results of the discussion with experienced industrial experts from a reputable aPS engineering company will be presented.

### A. Description of the Case Study

Part of a real industrial factory automation system for depalletizing trays was used for experimentation. Trays with parts (needles) are fed into the machine using conveyor belts, which are then depalletized by picking individual pieces off the tray using a 3-axis pick and place unit (PPU). Empty trays are subsequently moved out of the machine using a lift and conveyor system. A schematic view of the aPS is depicted in Fig. 5. The control software communicates with the hardware using 69 input variables and 26 output variables (mostly Booleans). It was written by the company participating in the evaluation workshop and comprises 119 POUs, adding up to about 15500 lines of code. The program uses two tasks, the first for the control logic (maximum scan cycle time of 10ms), and the second for the visualization (max. 100ms). It was implemented in IEC 61131-3 ST and SFC. The system is controlled by a Bosch Rexroth IndraControl VPP 21 Embedded PC with an Intel® Pentium® III 701 MHz processor and 504 MB of RAM. The control hardware uses the CODESYS Control RTE V3.5.5.20 real-time capable runtime and an Ethernet connection to connect to a development PC. The development system was used for generating and uploading the instrumented code and retrieving runtime information after test execution. A consumer laptop with an Intel® Core™ i7 5600U CPU running at 2.6 GHz and 8 GB RAM using Microsoft Windows 10 64-bit was used for this purpose. Here, CODESYS V3.5 SP8 Patch 1 and CODESYS Test Manager Version 4.0.1.0. implemented the approach together with the developed prototypical plug-in.

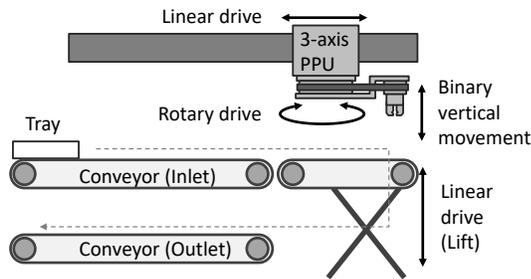

Fig. 5. Schematic for the industrial aPS for depalletizing trays using a 3-axis pick and place unit (PPU), which was used for the experiments [7]

*B. Experiments and Results*

A test suite was created for the system presented in the previous section based on a test plan provided by one of the industry partners. The test suite consists of 14 system test cases with a total runtime of about 25 minutes, testing the machine in different operation modes (manual and automatic) and in the case of an operation mode switch during automatic operation. All test cases include manual operations by the operator, such as putting a filled tray into the machine or acknowledging that the gripper is indeed closed. The test suite was created with the notion that most important functions in the machine were tested. The set of test cases was approved by an industrial expert from the company.

Using the test suite, two experiments were designed and performed. The goal for experiment I was to investigate the properties of the overhead generated by the different execution tracing implementations relating to the different prioritization methods. In experiment II, the prioritization approach itself was investigated for its benefits in aPS engineering.

*1) Experiment I: Runtime and Memory Overhead*

To acquire information about the runtime properties of the approach, two test cases were each executed five times for four different configurations of the control software. The four configurations represent different levels of the implementation of the approach, starting from 1. an uninstrumented, original control program and 2. an instrumented program implementing only the semi-automatic testing approach without any runtime information acquisition. Configuration 3. only allows for the unrefined prioritization of modification-traversing test cases, whereas 4. additionally allows for the usage of both refined prioritization methods described in this paper.

The two test cases chosen for the acquisition of runtime information were a test of a manual operation and one for automatic operation, as these are the most different regarding the involved code. During execution of the test cases, the average and maximum execution times of the PLC scan cycle, including reading sensor and writing actuator values, were measured using the task monitor of the IDE. From all measurements acquired during the execution of both test cases and all their repetitions (ten repetitions per test case), the average and maximum value were calculated. In addition, the required time to generate the dependency model and the required memory of the compiled program for each configuration were recorded.

The instrumentation of the project required less than one second, generating the dependency model and inserting 2261 trace function calls into the code. Each system test was executed completely without breaking real-time restrictions (10ms for the logic task), while all execution traces were written into the memory of the embedded execution hardware. The transfer from the memory to the hard drive of the embedded execution hardware was performed asynchronously after each test completion to avoid influencing real-time properties during test execution and did not exceed 10 PLC scan cycles, during which the program was still executed but in an idle state.

The required execution time for each of the configurations for the given application example (Fig. 5) is shown in Fig. 6 (left). While the average increase is quite significant, this increase has no direct influence on real-time requirements. For this, the maximum required execution time is relevant, which also increases, but to a smaller extent. Compared to the original program, the overall increase using the most complex tracing mechanism ("prioritization") results in 33.9% for the case study. With the given maximum scan cycle time of 10ms for this case study, all approaches are well within the bounds of real-time requirements, none exceeding 5ms. Therefore, all prioritization methods would be applicable to the given example (Fig. 5). Improvements to the runtime properties were implemented after the evaluation and are presented in Section VII.

Regarding the required memory for executable code on the execution hardware (see Fig. 6, right), the additionally needed space only increases moderately with about 23% for each of the prioritization methods. Yet, the increase in the needed global data increases significantly, resulting in an overall increase of about 136% for the full-featured prioritization method. The increase is mostly due to the prototypical implementation of the guided testing approach, accounting for the biggest increase. While a significant increase in the required memory was found, the overall required memory is still low with less than 5MB. Similar to the runtime overhead, improvements were made following the evaluation and are briefly presented in Section VII.

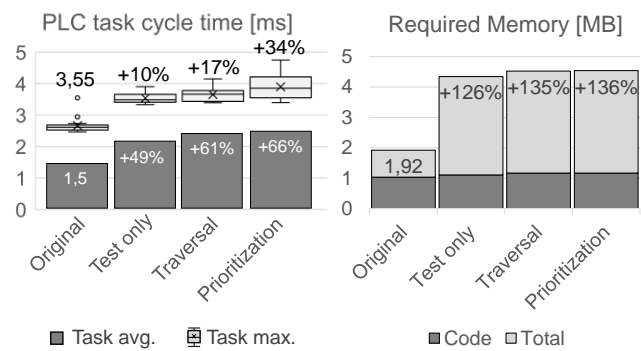

Fig. 6. Overheads of the approach: Required PLC scan cycle time for the different prioritization approaches (left) and required memory (right) in comparison to the original program.

*2) Experiment II: Software Change Scenario*

In the second experiment, a change scenario on a previously tested system was conducted to gain knowledge about the properties of the approach regarding prioritization. In this scenario, the timing of the gripper of the pick and place unit was adjusted to allow more consistent results regarding the identification of picked-up work pieces (needles). In sporadic cases, the gripper would not recognize a gripped needle even though it was holding onto one. The identification of gripped work pieces is achieved by gripping, waiting and then checking a vacuum sensor (Fig. 7, "_SnsNdl") that yields a different result in the case a work piece is present. The change relates to the waiting time, which was prolonged to allow for a more consistent buildup of a vacuum and thus a more consistent identification of gripped needles. For this, the assignment of a global variable "DelayNeedle" is adjusted that refers to the waiting time before checking the vacuum sensor.

A schematic view of the change scenario is depicted in Fig. 8. By changing the assignment of the global variable "DelayNeedle," several influenced parts of the program can be identified. The changed assignment renders the variable "DelayNeedle" modified. As it is used as an input for a call of the timer-POU instance "SqTimer", the called POU instance is also possibly affected by the change. Thus, the basic block (Fig. 7, "BB1") containing the timer is added to the set of modified basic blocks. The timer is used in two decisions (needle detected or not), possibly changing the progression of the program through the code (Fig. 7, "BB2" and "BB3").

This information is used to relate the modified parts of the code to the timing information acquired during the previous execution of the test cases, which is depicted in Table II. Not all test cases traverse all modifications: only the three test cases 11, 12 and 13 traverse some or all of the modifications. Thus, the basic prioritization adds these test cases to the group of high-priority test cases, whereas the rest of the test cases are assigned a low priority. For the refined prioritization, the order of these three test cases is calculated.

The refined prioritization method regarding intense traversal uses the traversal count of each test case to calculate

TABLE II
TIMING INFORMATION OF ALL SYSTEM TEST CASES FROM EXPERIMENT II REGARDING THE IDENTIFIED CHANGE AND CHANGE IMPACT

| System tests (Total execution time) | Basic block 1 traversal | Basic block 2 traversal | Basic block 3 traversal |
|---|---|---|---|
| 1.-10. Manual functions (14s-91s) | No traversal | No traversal | No traversal |
| 11. Empty tray (40s) | 5 times, first after 23s | No traversal | 5 times, first after 23s |
| 12. Partially filled tray (1m 33s) | 13 times, first after 25s | 8 times, first after 25s | 5 times, first after 52s |
| 13. Full tray (9m 47s) | 192 times, first after 24s | 192 times, first after 24s | No traversal |
| 14. Op-Mode-Change (3m 59s) | No traversal | No traversal | No traversal |

Test cases for manual functions are combined in this table, as none of these test cases traverses the modifications.

the number of times the test case interacts with a changed part of the control program. In this case, the result would be $p_{it}$=0.25 (modification traversals per second) for test case 11 ((5+5) traversals / 40s = 0.25 traversals per second), 0.28 for test case 12 and 0.65 for test case 13. Thus, the test case order would be 13, 12 and 11, followed by the rest of the test cases in no particular order. In test case 13, a full tray is depalletized, picking off 192 needles. Thus, the new timing is tested most by this test case, but not in all situations: there never is an empty spot on the tray, not testing whether an empty gripper is correctly recognized by the control program.

The calculation of the prioritization for quick-modification traversal returns several modification-traversing test combinations (MTTC):

MTTC 1: Test cases 11 + 12, resulting in a total time to traverse all modifications of 1m 5s (40s + 25s)

MTTC 2: Test cases 11 + 13, resulting in 1m 4s (40s + 24s)

MTTC 3: Test case 12, resulting in 52s

MTTC 4 & 5: Test cases 13 + 11 / 12, resulting in over 10m

Prioritizing the quickest combination, MTTC 3 is chosen, containing only a single test case (12). This test case uses a partially filled pallet, thus including two scenarios for the gripper: occupied and empty spaces on the tray. Therefore, this test case would test all modifications the quickest. Another close competitor would have been the combinations of an empty tray followed by a partially filled (MTTC 1) or full tray (MTTC 2).

*C. Discussion*

The results of the experiments and the performed measurements were presented to six experienced experts from a successful company in the field of aPS engineering. In the discussion, the satisfaction of the requirements of the approach was evaluated and will be presented, followed by the identified benefits of the approach.

*1) Expert discussion*

The group participating in the discussion of the experiment result comprised industrially experienced employees active in the fields of aPS commissioning, maintenance, software engineering and group management (technical development) from the company engineering the machine used in the case study. In addition to discussing the results in relation to the approach's applicability, a questionnaire was filled out by each expert to render the results more precisely.

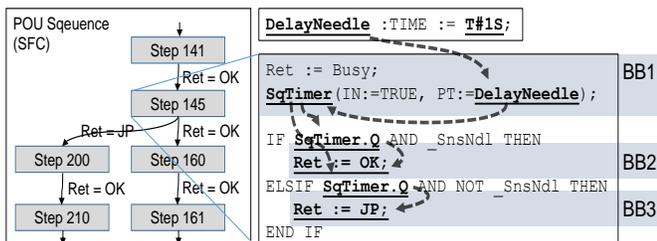

Fig. 7. Experiment II: Influence of a software change regarding a global variable assignment, possibly influencing the control flow through several Basic Blocks (BB).

TABLE III
RATING OF THE PRESENTED APPROACH BY A GROUP OF INDUSTRIAL EXPERTS REGARDING THE FULFILLMENT OF THE INITIAL REQUIREMENTS

| Industrial Requirement | Experts' Opinion |
|---|---|
| R1 Support of industrial software properties | Fulfilled – A representative industrial example was used in the case study |
| R2 Real time capability and memory size | Partially fulfilled – Execution time overhead suitable for about 20%-33% of the machines produced by the company, memory overhead acceptable for about 90% of the machines |
| R3 Inclusion of valid hardware and process behavior in the testing process | Fulfilled – Real hardware in combination with the software was used for system testing |
| R4 Manipulation of hardware and process behavior during testing | Fulfilled – Manual manipulation was part of the test cases |
| R5 Increase in efficiency during the testing process of changes | Fulfilled – Valuable support for pre-selection of test cases in case of changes was given by the approach |

The result of the rating of the approach by the group of experts is summarized in Table III. Regarding the requirements 1, 3 and 4, the experts approved the fulfillment of the requirements by the design and successful conduction of the case-study experiments. A representative application example with real industrial code was used and realistic test cases and change scenarios were executed.

Requirement 2 was seen as only partially fulfilled. In particular, the increase in required execution time was seen as critical for a larger segment of the machines produced by this company. Especially for highly automated machines with very short scan cycle times, the increase in maximum execution time would prevent the approach to be applied in its current state. It has to be noted that the current state of the algorithm was not fully optimized for speed and thus could be improved to increase the applicability of the approach.

The requirement regarding the increase in efficiency during the testing process of changes (R5) was seen as fulfilled. It was agreed upon that the approach would represent a valuable support in preselecting and prioritizing suitable test cases, which then could be further prioritized by the testing technician. This initial prioritization of test cases would save the involved personnel significant amounts of scarcely available time. While this property of the approach could not be quantified in the presented case study, the improvement of the regression testing process was seen as improved from a grade of 4.33 to 2.67 (1 = best, 7 = worst) on average.

*2) Benefits of the Approach*

The main benefit of the presented approach is the automatic support in prioritizing system test cases that are closely related to a change of the system. This prioritization is based on previous executions and an analysis of the change impact. Several problems are in focus by the presented approach:

1. Finding test cases that relate to a change of a system
2. Analyzing side effects of this change
3. Finding and prioritizing test cases that efficiently test possible regressions caused by side effects.

According to the industrial experts' opinion, the approach shows promising results in tackling these problems, improving the testing process of changes to aPSs regarding efficiency and testing quality. It is expected that when using the approach, regressions in systems can be found more quickly and reliably. At the same time, the industrial requirements were satisfied for a representative case study, and a discussion showed that the approach in its current prototypical form is already applicable for a significant part of the machines produced in this company and aPS engineering.

VII. POST-EVALUATION PERFORMANCE IMPROVEMENTS

Due to the identified shortcomings of the approach regarding real-time performance that were identified during the evaluation, several improvements were performed on the approach. These included (i) using inline tracing statements rather than using call-by-value function calls for each trace point, (ii) using structures instead of arrays for saving traces during runtime and (iii) the removal of unnecessary tracing functionality used for the debugging of the prototypical tool.

The resulting improvements of the approach regarding PLC scan cycle overhead are depicted in Fig. 8. The data was collected during the execution of twenty test case executions (ten repetitions of the two test cases from the evaluation in Section VI.B)). The increase in average scan cycle time (Fig. 8. bar chart, bottom) of the old approach of up to 66% could be reduced to an increase of 12% of required execution time in comparison to the original, unchanged program. Regarding the maximum required scan cycle times (Fig. 8. box chart, top), the longest observed scan cycle times of the new versions of the approach were shorter in comparison to the original program (Fig. 8. box chart, top, percentage values without parentheses). Within the interquartile range of the observed values, however (Fig. 8. box chart, top, percentage values in parentheses), an increase of about 15% of the required maximum execution time was observed.

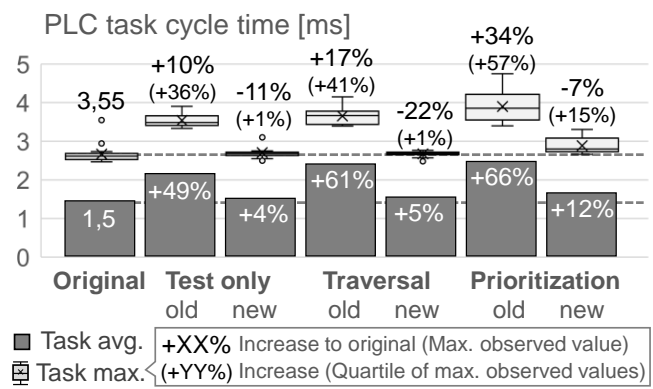

Fig. 8. Required maximum and average PLC scan cycle times for the different prioritization approaches (old version used in the evaluation and new, improved version) in comparison to the original program.

Regarding the required memory, the optimized approach only required up to 22% more memory compared to the 136% required in the old version. This was mostly due to the removal of unneeded tracing functionality used for debugging the prototypical implementation itself.

With the given improvements on the overhead characteristics of the approach, a higher applicability than identified in the evaluation would be possible.

## VIII. Conclusion and Outlook

A novel approach was presented that optimizes the regression testing process of fully integrated automated production systems (aPS) after modifications were performed on the control software. It supports test technicians in automatically finding and sequencing available test cases that are most likely to unveil newly introduced faults (regressions) based on runtime information acquired during previous executions, taking possible side effects of the modification into account. This prioritization helps to identify regressions to the system earlier, streamlining the regression testing process, which is often performed on-site under extremely tight time restrictions.

Through close cooperation with internationally renowned industrial partners, the approach was developed in line with relevant industrial requirements. Particularly, the approach is independent of simulations and regards real-time properties relevant for aPSs, which hinder the industrial applicability of most other approaches proposed in literature. Through the application of the approach in an industrial case study and a subsequent evaluation with experienced industrial experts, it was found that the approach shows very promising results regarding support during the regression testing process of aPSs and industrial applicability.

For further research, different properties of the approach could be improved to allow for an even wider applicability. Especially the runtime overhead could be decreased even further. From the first investigations, the overhead is related to both the cyclomatic complexity of the code (more complexity requires more trace function calls) and the amount of instructions in basic blocks (the more instructions, the lower the percentage of overhead from the trace functions). Yet, the maximum overhead seems highly dependent on task scheduling and the path that was taken through the code, where worst-case execution time analysis [35] could be an appropriate means to determine possible real-time breaches. Certainly, more research on this property of the approach would be an interesting focus of future work. A combination with other efficient tracing techniques, such as [29], could also increase applicability for aPSs with very short scan cycles. Furthermore, the support of all IEC 61131-3 programming languages is of interest in making the approach applicable for all control programs programmed in this standard. The change impact analysis and prioritization algorithms should also be a focus of further research, improving the safety of the choice of test cases to allow for the omission of low-priority test cases without compromising the quality of the testing process.

## Acknowledgment

The authors would like to express their gratitude to the companies 3S – Smart Software Solutions GmbH, Bosch Rexroth AG and Robert Bosch GmbH for their valuable input and support during the development and evaluation of the presented approach during and after the project MoBaTeSt, funded by the Bavarian Ministry of Economic Affairs and Media, Energy and Technology (Grant IUK413).

## References

[1] B. Vogel-Heuser, A. Fay, I. Schaefer, and M. Tichy, "Evolution of software in automated production systems: Challenges and research directions," *J. Syst. Softw.*, vol. 110, pp. 54–84, Dec. 2015.

[2] V. Vyatkin, "Software Engineering in Industrial Automation: State-of-the-Art Review," *IEEE Trans. Ind. Informatics*, vol. 9, no. 3, pp. 1234–1249, Aug. 2013.

[3] M. L. Alvarez, I. Sarachaga, A. Burgos, E. Estevez, and M. Marcos, "A Methodological Approach to Model-Driven Design and Development of Automation Systems," *IEEE Trans. Autom. Sci. Eng.*, pp. 1–13, 2016.

[4] R. Hametner, A. Zoitl, and M. Semo, "Automation component architecture for the efficient development of industrial automation systems," in *IEEE International Conference on Automation Science and Engineering (CASE)*, 2010, pp. 156–161.

[5] F. Basile, P. Chiacchio, and D. Gerbasio, "On the Implementation of Industrial Automation Systems Based on PLC," *IEEE Trans. Autom. Sci. Eng.*, vol. 10, no. 4, pp. 990–1003, Oct. 2013.

[6] IEC, "IEC 61131 Programmable Controllers - Part 3: Programming Languages (Second Edition)." International Electrotechnical Commission Std., 2003.

[7] S. Ulewicz and B. Vogel-Heuser, "Guided semi-automatic system testing in factory automation," in *IEEE International Conference on Industrial Informatics (INDIN)*, 2016, pp. 142–147.

[8] S. Ulewicz and B. Vogel-Heuser, "System regression test prioritization in factory automation: Relating functional system tests to the tested code using field data," in *Annual Conference of the IEEE Industrial Electronics Society (IECON)*, 2016, pp. 4619–4626.

[9] H. Prähofer, F. Angerer, R. Ramler, H. Lacheiner, and F. Grillenberger, "Opportunities and challenges of static code analysis of IEC 61131-3 programs," in *IEEE International Conference on Emerging Technologies and Factory Automation (ETFA)*, 2012.

[10] E. Estevez and M. Marcos, "Model-Based Validation of Industrial Control Systems," *IEEE Trans. Ind. Informatics*, vol. 8, no. 2, pp. 302–310, May 2012.

[11] S. Biallas, J. Brauer, and S. Kowalewski, "Arcade.PLC: A verification platform for programmable logic controllers," in *IEEE International Conference on Automation Science and Engineering (CASE)*, 2012, pp. 338–341.

[12] O. Ljungkrantz, K. Åkesson, M. Fabian, and Chengyin Yuan, "Formal Specification and Verification of Industrial Control Logic Components," *IEEE Trans. Autom. Sci. Eng.*, vol. 7, no. 3, pp. 538–548, Jul. 2010.

[13] S. Ulewicz, M. Ulbrich, A. Weigl, M. Kirsten, F. Wiebe, B. Beckert, and B. Vogel-Heuser, "A verification-supported evolution approach to assist software application engineers in industrial factory automation," in *IEEE International Symposium on Assembly and Manufacturing (ISAM)*, 2016, pp. 19–25.

[14] S. Rösch, S. Ulewicz, J. Provost, and B. Vogel-Heuser, "Review of Model-Based Testing Approaches in Production Automation and Adjacent Domains—Current Challenges and Research Gaps," *J. Softw. Eng. Appl.*, vol. 8, no. 9, pp. 499–519, 2015.

[15] R. Hametner, B. Kormann, B. Vogel-Heuser, D. Winkler, and A. Zoitl, "Test case generation approach for industrial automation systems," in *International Conference on Automation, Robotics and Applications (ICARA)*, 2011, pp. 57–62.

[16] S. Rösch and B. Vogel-Heuser, "A light-weight fault injection approach to test automated production system PLC software in industrial practice," *Control Eng. Pract.*, vol. 58, no. March 2016, pp. 12–23, Jan. 2017.

[17] S. Süß, S. Magnus, M. Thron, H. Zipper, U. Odefey, V. Fassler, A. Strahilov, A. Klodowski, T. Bar, and C. Diedrich, "Test methodology for virtual commissioning based on behaviour simulation of production

systems," in *IEEE International Conference on Emerging Technologies and Factory Automation (ETFA)*, 2016, vol. 2016–Novem, pp. 1–9.

[18] P. Puntel-Schmidt and A. Fay, "Levels of Detail and Appropriate Model Types for Virtual Commissioning in Manufacturing Engineering," *IFAC-PapersOnLine*, vol. 48, no. 1, pp. 922–927, 2015.

[19] M. Barth and A. Fay, "Automated generation of simulation models for control code tests," *Control Eng. Pract.*, vol. 21, no. 2, pp. 218–230, Feb. 2013.

[20] S. Yoo and M. Harman, "Regression testing minimization, selection and prioritization: a survey," *Softw. Testing, Verif. Reliab.*, vol. 22, no. 2, pp. 67–120, 2012.

[21] M. D. Ernst, "Static and dynamic analysis: synergy and duality," *ICSE Work. Dyn. Anal.*, pp. 24–27, 2003.

[22] IBM, "Rational DOORS," 2016. [Online]. Available: http://www-03.ibm.com/software/products/en/ratidoor. [Accessed: 05-May-2017].

[23] A. Zeller and M. Weyrich, "Test case selection for networked production systems," in *IEEE International Conference on Emerging Technologies and Factory Automation (ETFA)*, 2015.

[24] P. Caliebe, T. Herpel, and R. German, "Dependency-based test case selection and prioritization in embedded systems," in *IEEE International Conference on Software Testing, Verification and Validation (ICST)*, 2012, pp. 731–735.

[25] H. Baller, S. Lity, M. Lochau, and I. Schaefer, "Multi-objective test suite optimization for incremental product family testing," in *IEEE International Conference on Software Testing, Verification and Validation (ICST)*, 2014, pp. 303–312.

[26] S. Ulewicz, D. Schütz, and B. Vogel-Heuser, "Software changes in factory automation: Towards automatic change based regression testing," in *Annual Conference of the IEEE Industrial Electronics Society (IECON)*, 2014, pp. 2617–2623.

[27] A. Orso, T. Apiwattanapong, and M. J. Harrold, "Leveraging field data for impact analysis and regression testing," *ACM SIGSOFT Softw. Eng. Notes*, vol. 28, no. 5, p. 128, Sep. 2003.

[28] G. Rothermel, R. H. Untch, Chengyun Chu, and M. J. Harrold, "Prioritizing test cases for regression testing," *IEEE Trans. Softw. Eng.*, vol. 27, no. 10, pp. 929–948, 2001.

[29] H. Prähofer, R. Schatz, C. Wirth, and H. Mössenböck, "A comprehensive solution for deterministic replay debugging of SoftPLC Applications," *IEEE Trans. Ind. Informatics*, vol. 7, no. 4, pp. 641–651, 2011.

[30] ISO/IEC/IEEE, "ISO/IEC/IEEE 24765:2010 Systems and Software Engineering - Vocabulary." ISO/IEC/IEEE, 2010.

[31] S. Feldmann, F. Hauer, S. Ulewicz, and B. Vogel-Heuser, "Analysis framework for evaluating PLC software: An application of Semantic Web technologies," in *IEEE International Symposium on Industrial Electronics (ISIE)*, 2016, pp. 1048–1054.

[32] 3S - Smart Software Solutions GmbH, "CODESYS Development System," 2016. [Online]. Available: https://www.codesys.com/products/codesys-engineering/development-system.html. [Accessed: 05-May-2017].

[33] 3S - Smart Software Solutions GmbH, "CODESYS Test Manager," 2016. [Online]. Available: http://store.codesys.com/codesys-test-manager.html. [Accessed: 05-May-2017].

[34] P. Runeson, M. Höst, A. Rainer, and B. Regnell, *Case Study Research in Software Engineering*. Hoboken, NJ, USA: John Wiley & Sons, Inc., 2012.

[35] R. Wilhelm, T. Mitra, F. Mueller, I. Puaut, P. Puschner, J. Staschulat, P. Stenström, J. Engblom, A. Ermedahl, N. Holsti, S. Thesing, D. Whalley, G. Bernat, C. Ferdinand, and R. Heckmann, "The worst-case execution-time problem—overview of methods and survey of tools," *ACM Trans. Embed. Comput. Syst.*, vol. 7, no. 3, pp. 1–53, Apr. 2008.

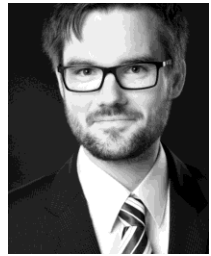

**Sebastian Ulewicz** received a diploma engineer's degree from the Technical University of Munich, Germany, in 2012. He is currently pursuing a Ph.D. degree with the Institute of Automation and Information Systems, Technical University of Munich, Germany. His current research interests include the model-based testing of automated production systems, regression testing and test coverage analysis.

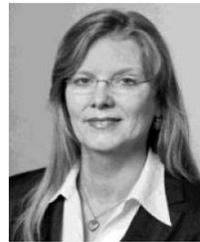

**Birgit Vogel-Heuser** received a Dr.-Ing. Degree in electrical engineering and a Ph.D. degree in mechanical engineering from RWTH Aachen University, Aachen, Germany, in 1991. She was involved in industrial automation with the machine and plant manufacturing industry for nearly ten years. After holding different chairs of automation in Hagen, Wuppertal, and Kassel, she has been the Head of the Automation and Information Systems Institute with the Technical University of Munich, Munich, Germany, since 2009. Her current research interests include systems and software engineering, and the modeling of distributed and reliable embedded systems. Prof. Vogel-Heuser is the Coordinator of the Collaborative Research Centre SFB 768: Managing Cycles in Innovation Processes—Integrated Development of Product-Service Systems Based on Technical Products.